\documentclass[preprint,12pt,sort&compress]{elsarticle}

 \usepackage{graphicx}

\usepackage{amssymb}
\usepackage{amsmath}

\usepackage{color}
\usepackage{epstopdf}
\usepackage{upgreek}
\usepackage{url}
\usepackage{siunitx}
\usepackage{algorithm}
\usepackage[noend]{algpseudocode}

\usepackage{float}

\usepackage{url}
\usepackage{hyperref}

\usepackage{xspace}
\newcommand{\figurerefname}{Figure}
\newcommand{\tablerefname}{Table}

\newcommand{\fref}[1]{\figurerefname~\ref{#1}\xspace}
\newcommand{\tref}[1]{\tablerefname~\ref{#1}\xspace}

\journal{npj Computational Materials}

\begin{document}

\begin{frontmatter}



\author[1]{Markus Gusenbauer\corref{email}}
\author[1]{Harald Oezelt}
\author[1]{Johann Fischbacher}
\author[1]{Alexander Kovacs}

\author[2]{Panpan Zhao}
\author[2]{Thomas George Woodcock}
\author[1]{Thomas Schrefl}

\address[1]{Department for Integrated Sensor Systems, Danube University Krems, Austria}
\address[2]{Leibniz IFW Dresden, Institute for Metallic Materials, Germany}
\cortext[email]{Corresponding author: \texttt{markus.gusenbauer@donau-uni.ac.at}}

\title{Extracting local switching fields in permanent magnets using machine learning}

\begin{abstract}
Microstructural features play an important role for the quality of permanent magnets. The coercivity is greatly influenced by crystallographic defects, which is well known for MnAl-C, for example. In this work we show a direct link of microstructural features to the local coercivity of MnAl-C grains by machine learning. A large number of micromagnetic simulations is performed directly from Electron Backscatter Diffraction (EBSD) data using an automated meshing, modeling and simulation procedure. Decision trees are trained with the simulation results and predict local switching fields from new microscopic data within seconds. 
\end{abstract}

\end{frontmatter}


\section{\label{sec:intro}Introduction}

Permanent magnets are of great interest in today's economy. Especially green energy applications such as in wind turbines and hybrid/electric vehicles demand high performance permanent magnets.  The performance of permanent magnets is mainly determined by intrinsic magnetic properties and microstructural features. The intrinsic properties are adjusted by including rare earth (RE) elements which mainly increase the magnetocrystalline anisotropy. Due to economical and environmental considerations there is a high interest in reducing the use of RE elements~\cite{skokov2018heavy}. MnAl-C contains no RE or other critical raw materials and its intrinsic magnetic properties make it an attractive alternative to certain types of RE-based permanent magnets~\cite{coey2012permanent}. The microstructure of MnAl-C magnets is known to contain a range of defects, such as grain boundaries, twins and antiphase boundaries~\cite{landuyt1978defect,houseman1983domain,yanar2001evolution,bittner2015twin,palanisamy2019compositional}.

Using micromagnetic simulations the influence of such microstructures on the performance of the magnet was analyzed~\cite{bance2017role}. Recently we developed a workflow to directly compute the coercive field from selections of EBSD data~\cite{gusenbauer2019automated}. The resolution of the microscopic images and the needed scaling for the micromagnetic simulations remain  key challenges in analyzing bulk permanent magnets~\cite{fischbacher2018micromagnetics}. The acquisition of EBSD data sets encompassing many hundreds of $\upmu$m is routinely possible, but current micromagnetic simulation methods are still limited to a size range of only a few $\upmu$m.

A promising approach to reduce the computational costs is the use of machine learning to predict the coercivity of permanent magnets~\cite{exl2018magnetic}. Machine learning has been already used for the characterization of steel microstructures~\cite{azimi2018advanced}, for example. In Mg-alloys, Orme and co-workers analyzed the formation and propagation of twinning boundaries using decision trees~\cite{orme2016insights}.

One step further is the direct link of microstructural features to the coercivity of permanent magnets. Kronm\"uller and Goll discussed microstructural properties of advanced hard magnetic materials and its relation to the quality of permanent magnets~\cite{kronmuller2009micromagnetism} . By investigating the various boundary types of MnAl-C, their distribution and the effect on the coercivity by micromagnetic simulations, we hope to optimize the overall performance of MnAl-C magnets. In this work we use trained regression trees to map microstructural features with high accuracy to local coercivities of spatially confined areas. The training data is obtained directly from EBSD data of MnAl-C. Using automated geometry construction, discretization of the EBSD map and micromagnetic demagnetization simulations, a large amount of training data can be obtained. We identified the most prominent microstructural feature combinations to accurately predict the coercivity.

\section{\label{sec:methods}Methods and Theory}

In the following section we show an automated modeling and meshing procedure to generate a large amount of simulation data from real EBSD maps. Micromagnetic simulations are performed to compute the nonreversible switching fields of spatially confined areas. We show the important microstructural attributes of the EBSD maps for machine learning as well es the used regressional decision trees. 

\subsection{\label{sec:meth_data}Dataset generation}

\begin{figure}[H]
\begin{center}
 \includegraphics[width=0.8\textwidth]{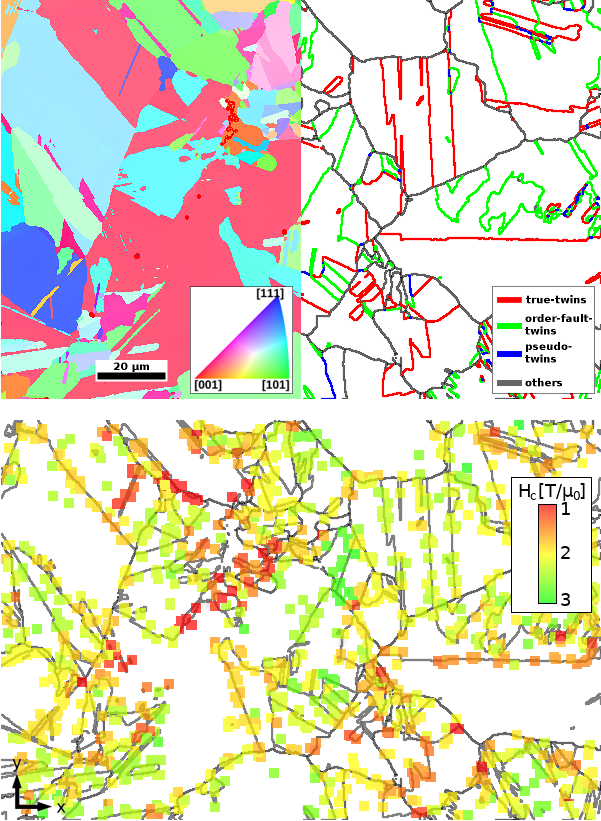}
\end{center}
 \caption{EBSD map of 600x400 pixels with a pixel edge length of 0.3 $\upmu$m divided in two halves (top). Inverse pole figure (IPF) map converted with
Dream3D (dream3d.bluequartz.net, last visited on 17/12/2019) on the left. Crystallographic twin boundaries on the right: true-twins (red), order-fault-twins (green), pseudo-
twins (blue) and others (gray). Each colored square (10x10 pixels) corresponds to the switching field $H_\mathrm{c}$, computed by micromagnetic simulations (bottom).}
 \label{ebsd}
\end{figure}

We are computing the switching fields $H_\mathrm{c}$ of MnAl-C selections obtained from EBSD data. The original data is represented as crystallographic orientations on a regular grid. Our EBSD sample in \fref{ebsd} has a size of 180x120 $\upmu$m$^2$ (600x400 pixels). We randomly pick 10x10 pixels from the original dataset to create finite element meshes for the simulations. In order to avoid a biased dataset and the recomputation of equal pixel frames the uniqueness is checked in advance. \fref{ebsd} on the bottom shows the grain boundaries of the EBSD map and the random simulation selections (about 1500) colored by its computed coercivity. Due to the uniqueness constraint, the selections are preferentially located at the grain boundaries. Since nucleation usually starts near defects, which for MnAl-C are most often grain or twin boundaries (\fref{ebsd} on the top right), those locations are the most important determining factors for the bulk coercivity.  

\begin{figure}[H]
\begin{center}
 \includegraphics[width=0.9\textwidth]{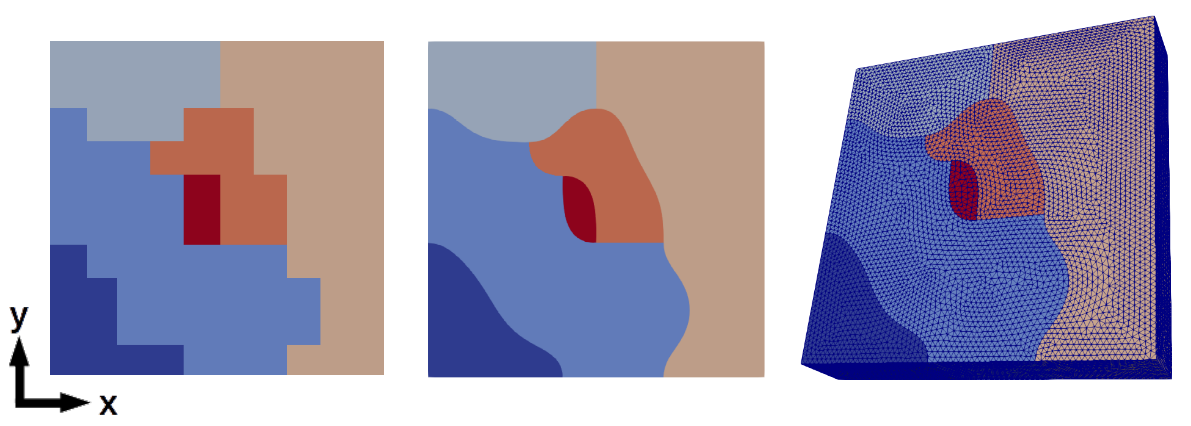}
\end{center}
 \caption{Original EBSD data selection (left), smoothed grain boundaries (center) and tetrahedral finite element mesh (right) with arbitrary colors according to the grain id. Simulation parameters are given in \tref{parameters}.}
 \label{smooth}
\end{figure}

The resolution of the EBSD dataset is not high enough to directly compute micromagnetic simulations. Features like grain boundaries which are small compared to the grid resolution are not accurately spatially resolved. Sharp angles need to be smoothed for the simulations in order to reduce numerical instabilities. In previous work we showed an automated meshing procedure for the fast generation of simulation models~\cite{gusenbauer2019automated}. We combine multiple steps of the toolchain into a single meshing tool using the pre- and post-processing software Salome (salome-platform.org, last visited on 29/07/2019). We load the pixelated EBSD data into the software, smooth the grain boundaries and create a high quality finite element mesh (Algorithm~\ref{alg}). The software basically finds all intersection points of the grain boundaries and reconnects them with Bezier curves, replacing the stepped lines of the original pixelated EBSD image. \fref{smooth} shows an example of an original EBSD selection on the left, the creation of a smoothed 2-dimensional surface in the center and the extruded 3-dimensional mesh on the right.

The mesh is then used by a fast micromagnetic solver, which computes the switching field for the particular selection.

\begin{algorithm}
\caption{Pseudo code for automatically generating high quality finite element meshes from EBSD data using the Salome Python interface (salome-platform.org, last visited on 29/07/2019)}\label{alg}
\begin{algorithmic}[1]
\Require EBSD 2-dimensional dataset with \textit{Ids} for each grain (\textit{data.txt}) and orientation information of each grain \textit{Id} (\textit{orientations.txt}) 
\State Create a \textit{Face} for each pixel at its particular position and merge equal Pixel/Face \textit{Ids} to \textit{Partitions}
\Ensure Each individual grain is represented as \textit{Partition}
\State Find \textit{Closed\_Free\_Boundaries} for each grain \textit{Partition}
\Ensure Each grain \textit{Partition} has a boundary wire
\State Find \textit{Common\_Wires} of all \textit{Closed\_Free\_Boundaries}
\Ensure All shared boundaries between individual grains found, consequently all intersection points of the boundary wires are found
\State Extract and sort \textit{Vertices} from each \textit{Common\_Wire}
\Ensure \textit{Vertices} are in the correct order for redrawing
\State Create \textit{Bezier\_Curve} from each list of \textit{Vertices}
\Ensure All \textit{Bezier\_Curves} are valid, otherwise split erroneous \textit{Common\_Wires} to reduce length, and redo until valid
\State Create \textit{Partition} of all \textit{Bezier\_Curves} with a rectangular \textit{Face} with the size of the EBSD selection
\Ensure Curves and face \textit{Partition} is valid, otherwise reduce \textit{LimitTolerance}
\State Extract new grain \textit{Faces} of curves and face \textit{Partition} and extrude in the 3rd dimension to obtain grain \textit{Solids}
\State Create and export final \textit{Mesh} using original grain \textit{Id} from \textit{data.txt} and orientation information from \textit{orientations.txt}
\end{algorithmic}
\end{algorithm}
\subsection{\label{sec:meth_magn}Micromagnetic simulations}
The finite element mesh obtained from the automated meshing routine is used by a fast micromagnetic solver. We are computing magnetization reversal curves using the Landau-Lifshitz-Gilbert (LLG) equation~\cite{suess2002time}. Simulation parameters listed in \tref{parameters} are obtained from bulk MnAl-C~\cite{THIELSCH201725} at 300 K. The largest edge length of the tetrahedral finite elements (mesh size) is set to be smaller than the smallest characteristic length of the material (compare findings in~\cite{gusenbauer2019automated}). Bance et al. downsized the original EBSD dataset to reduce computational costs by 1:500 and 1:50~\cite{bance2017role}. Here we use a scaling ratio of 1:15, so that 1 $\upmu$m in the original EBSD dataset corresponds to 67 nm in the simulations. 

\begin{table}[h]
\begin{center}
\begin{tabular}{ | l | r | l | }
  \hline
  \bf{Parameter} & \bf{Value} & \bf{Unit} \\
  \hline
  temperature $T$& 300 & K \\
  \hline                       
  magnetic polarization $J_\mathrm{s}$&0.8& T\\
  \hline
  exchange constant $A$&19.9&pJ/m\\
  \hline
  uniaxial anisotropy constant $K$& 1.5 &MJ/m$^3$\\
  \hline
  mesh size & 3 & nm\\
  \hline
  scaling ratio & 1:15 & \\
  \hline
  selection size & 10x10 & pixels\\
  \hline
  extrusion thickness & 2 & pixels\\
  \hline
\end{tabular}
\end{center}
 \caption{Micromagnetic parameters for the simulations. Parameters are obtained from bulk MnAl-C experiments~\cite{THIELSCH201725}}
\label{parameters}
\end{table}

The nonreversible switching fields $H_\mathrm{c}$ obtained by the micromagnetic simulations are the solutions (labels) for the machine learning dataset. We compute the hysteresis curves with an external field aligned in-plane from +4T to -4T. In \fref{ebsd} on the bottom we highlight $H_\mathrm{c}$ from red (1 T/$\upmu_0$) to green (3 T/$\upmu_0$). In some cases $H_\mathrm{c}$ is ambiguous. We discuss the problem in Section \ref{sec:res_labeling}.    

\subsection{\label{sec:meth_microstructure}Microstructural attributes}

From the EBSD data we have the orientation information of each grain of the simulation model. We can use each individual data point (pixel) of an EBSD selection as a feature or we can just use a list of single grains with their respective orientation. From the first we get also location coordinates, which the second does not include. We define the crystallographic orientation as unit vector pointing into the positive field direction y. The smoothing operation creates many more data points than the original image shows. A higher number of data points slows down the learning process, but does not significantly improve the machine learning accuracy. Therefore we stick to the discretization points of the original EBSD data for training. The size of the grains can effect the switching field as well and is therefore added to the list of microstructural features.

By using the Stoner-Wohlfarth (SW) model for single-domain ferromagnets~\cite{stoner1948mechanism} we analytically calculate a switching field approximation for each individual grain~\cite{kronmuller2009micromagnetism}:

\begin{equation}
H_\mathrm{c}=2 K/J_\mathrm{s}\left(\cos(\beta)^{2/3}+\sin(\beta)^{2/3}\right)^{-3/2},
\label{StWo}
\end{equation}

with the misorientation angle $\beta=\mathrm{acos}(\mathrm{e}_\mathrm{y})$ between the easy axis of a grain and the external field (unit vector $\mathrm{e}_\mathrm{y}$). The lowest field value of all grains gives the nonreversible switching event of the selection, if other effects like stray field and grain boundary exchange coupling  are neglected.

Twin boundaries are known to have a strong influence on the quality of permanent magnets. Especially in MnAl-C true-twins, order-fault-twins and pseudo-twins are frequently found~\cite{bittner2015twin}, which have a clearly defined misorientation angle. In previous work we computed hysteresis curves for selections of MnAl-C and showed a significant effect of twinning boundaries on the coercive field~\cite{gusenbauer2019automated,bance2017role}.  From manual observation of the simulation results and from partial dependency curves in Section~\ref{sec:res_features} we observe a negative influence on the coercivity, if the adjacent grains orientations have a large enclosed angle. Therefore we incorporate the twin angles into a feature for machine learning by calculating the minimum dot product of adjacent grains orientation unit vectors. A large angle implicitly means that both of the adjacent grains are disadvantageously orientated to the applied field direction. Whereas the SW model incorporates only the misorientation angle of a single grains orientation with respect to the external field.  

From the microstructural attributes and analytical equations we acquire the following features used for machine learning. In brackets the short feature name is given as well as the number of features as seen by the machine. 

\begin{itemize}
 \item Crystallographic orientation for each data point as unit vector (``PIXEL'', 300)
 \item Crystallographic orientation for each grain as unit vector (``GRAIN'', 60)
 \item Size of grains (``SIZE'', 20)
 \item Minimum Stoner-Wohlfarth switching field (``SW'', 1)
 \item Minimum dot product of adjacent grains orientation (``TWIN'', 1)
\end{itemize}

PIXEL's features are all orientation unit vectors of the data points (10x10 pixels). The maximum possible number of grains for our selections is set to 20. Therefore SIZE has 20 entries and GRAIN 20 times 3 components of the orientation vectors. If there are less than 20 grains in the selection the feature is set to 0. Typical feature correlation or feature elimination methods cannot be used in such a configuration, because each pixel is already a feature, hence cannot be reasonably compared to other features. In machine learning usually feature scaling is necessary. For our selected features the number range is very small, therefore scaling does not beneficially improve the outcome. Feature importance will be discussed in Section~\ref{sec:res_features}.

\subsection{\label{sec:meth_machlearn}Machine learning methods}

With our automated meshing procedure we are able to generate a large number of simulation models. Yet calculating the switching fields for each model comes with a certain computational cost. A single simulation on a single CPU takes a few hours depending on the magnetic configuration. The computation of $H_\mathrm{c}$ of all selections in \fref{ebsd} on the bottom requires about half a year of CPU time. By using a machine learning code we can reduce the time to solution for any new EBSD data to a few seconds. The training and test set contains microstructural features of the selected models and its switching fields (supervised learning). We are using Python with the Scikit-Learn framework~\cite{scikit-learn,geron2017hands}. Similar to Exl and co-workers~\cite{exl2018magnetic} we are using the two decision trees Random Forest (RF) and Gradient Boosting (GB). We tested convolutional networks with the pixelated EBSD data but could not achieve the same prediction accuracy. For simple comparison of labeling methods in Section~\ref{sec:res_labeling} we are using a classification of ``good'' and ``bad'' selections, which translates to a high and low switching field respectively. But our main focus is on the use of regression, because we want to predict actual values of the switching fields in a new EBSD dataset. For our machine learning approaches we use 80\% of the simulation models for learning and 20\% for testing.  

We are using the GridSearchCV method from Scikit-Learn to optimize the hyperparameters of our RF and GB regressors. In our case the optimized hyperparameters are

\begin{verbatim}
RandomForestRegressor(n_estimators=300, max_features=auto, 
   max_depth=100,min_samples_leaf=1, min_samples_split=2)
GradientBoostingRegressor(n_estimators=300, max_features='auto',
   max_depth=4, learning_rate=0.07, min_samples_leaf=5, 
   min_samples_split=8) 
\end{verbatim}

A Voting Regressor (VR) is used to average the individual predictions on the whole dataset and to form a final prediction. For quantitative analysis of the predicted values we are using the Mean Absolute Error (MAE).

\section{\label{sec:results}Results}

In the Results section we show the importance of accurate labeling. Nonreversible switching fields are sometimes difficult to determine. Afterwards we analyze the effect of the microstructural attributes as well as its combinations on the machine learning efficiency. The trained machine can be used to predict local coercivities of new microscopic data, even for different microscopic resolutions.

\subsection{\label{sec:res_labeling}Labeling}

\begin{figure}[H]
\begin{center}
 \includegraphics[width=0.6\textwidth]{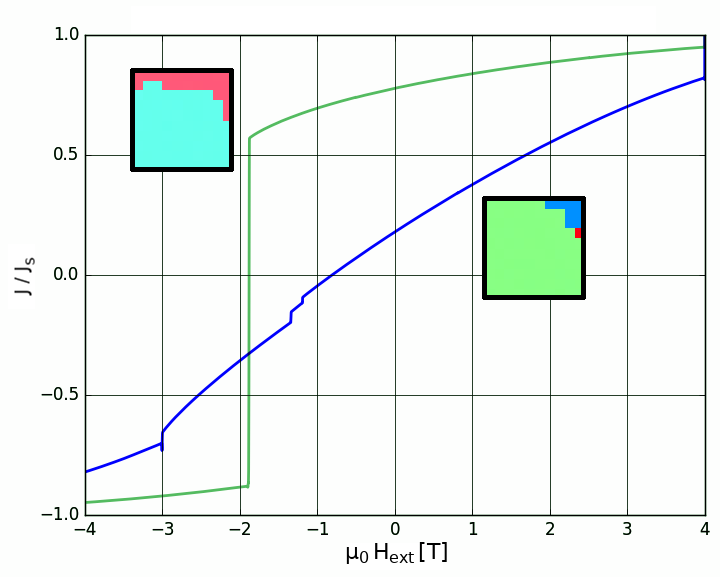}
\end{center}
 \caption{Exemplary hysteresis curves with clear (green) and uncertain (blue) definition for the nonreversible switching field. }
 \label{labeling}
\end{figure}

Training the machine requires a well defined set of microstructural features as well as an explicit solution, the nonreversible switching field $H_\mathrm{c}$. Often $H_\mathrm{c}$ is sufficiently defined by the field which is needed to reverse the sign of $J_\mathrm{s}$ as for example the green line in \fref{labeling}. For such a curve the magnet is clearly in a nonreversible state after the step. Whereas the blue line is much flatter and shows only three small steps, non of which is a clear candidate for $H_\mathrm{c}$.

In order to find the optimum label we created a simple classification problem, which tries to predict the 20\% percentile of the ``bad'' selections, with the lowest switching field. We tested 3 different definitions of the switching field $H_\mathrm{c}$, (a) the first step, (b) the last step and (c) the largest step of the curve. The initial step at 4T depends on the magnetic configuration of the selection and is not used to find the best label. We obtained prediction accuracies of 90\%, 85\% and 86\% respectively, therefore we decided to use the first step of the curve as definition of $H_\mathrm{c}$.

\subsection{\label{sec:res_features}Feature selection}
In Section \ref{sec:meth_microstructure} we defined 5 features from our microstructural EBSD dataset. We predicted the local coercivity via a single feature and summarized the MAE scores in \tref{res:singlefeatures}. The lowest error is reached with the feature GRAIN, in which the location information of the crystals is not included. The PIXEL feature, which contains the additional position information, performs a little worse. Probably, redundant data and the high number of features make the training more difficult. The features SW and TWIN produce a MAE of 0.185 T and 0.17 T respectively. SIZE can only be used in addition to other features, due to the large error.

\begin{table}[h]
\begin{center}
\begin{tabular}{ | l | c | c | c | c | c | c |}
  \hline
  & \bf{PIXEL} & \bf{GRAIN} & \bf{SIZE} & \bf{SW} & \bf{TWIN}\\
  \hline  
  MAE [T] & 0.151 & 0.136 & 0.257 & 0.185 & 0.170 \\
  \hline

\end{tabular}
\end{center}
 \caption{Single feature predictions: MAE value obtained by the Voting Regressor (VR).}
\label{res:singlefeatures}
\end{table}

\begin{figure}[H]
\begin{center}
 \includegraphics[width=0.7\textwidth]{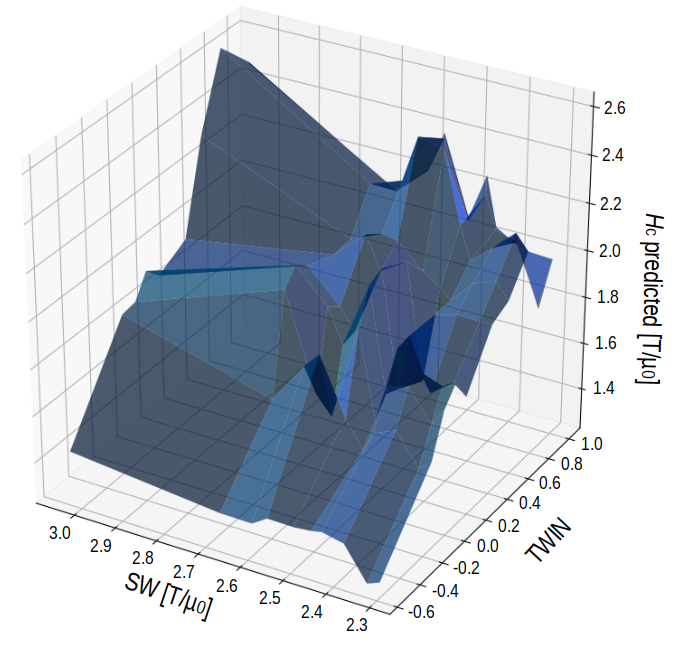}
\end{center}
 \caption{Two-way partial dependency surface with SW and TWIN.   }
 \label{partial_dependency}
\end{figure}

A two-way partial dependency surface of SW and TWIN shows the dependency on the predicted coercivity (\fref{partial_dependency}). The coercivity is definitely reduced by a negative dot product of two adjacent grains orientations (TWIN) even for a large SW. On the contrary SW is much closer to the predicted switching field $H_\mathrm{c}$ for a positive dot product. A clear maximum is located at a large SW and a large TWIN. Yet also several maxima can be found even for lower SW. Partial dependency curves for PIXEL, GRAIN and SIZE are not reasonable, as each one of it contains multiple features (data points in PIXEL, maximum number of possible grains in a selection for GRAIN and SIZE). 

The combination of several features introduces more possibilities for accurately predicting the coercivity. The best feature combinations are shown in \tref{res:bestfeatures}. Given the fact, that individual features or feature pairs may perform poorly, the best MAE score is still reached with a combination of all features. 

\begin{table}[H]
\begin{center}
\begin{tabular}{| c | c | c | c | c | c |}
  \hline
 \bf{PIXEL} & \bf{GRAIN} & \bf{SIZE} & \bf{SW} & \bf{TWIN} & \bf{MAE [T]}\\
    \hline 
  - & X & X & - & - & 0.120\\
     \hline 
  X & X & - & - & - & 0.119\\
     \hline 
  X & X & X & - & - & 0.119\\
     \hline 
  X & - & X & X & - & 0.118\\
    \hline 
  - & X & X & X & - & 0.118\\
    \hline 
  - & X & X & - & X & 0.118\\
    \hline 
  - & X & X & X & X & 0.114\\
    \hline 
  X & X & - & X & - & 0.113\\
    \hline 
  X & X & X & - & X & 0.112\\
    \hline 
  X & X & - & - & X & 0.112\\
  \hline 
  X & X & X & X & - & 0.110\\
  \hline  
  X & - & - & X & X & 0.108\\
  \hline 
  X & - & X & X & X & 0.105\\
  \hline
  X & X & - & X & X & 0.103\\
  \hline 
  X & X & X & X & X & 0.103\\
  \hline  
\end{tabular}
\end{center}
 \caption{Best feature combinations (MAE $\leq$ 0.12 T): MAE value obtained by the Voting Regressor (VR).}
\label{res:bestfeatures}
\end{table}

\subsection{\label{sec:res_complex}Microstructural complexity}

Right now we have used all the simulations for training and testing. For better understanding of the MAE values, we can split the data according to the number of grains in the selections. Each dataset with the same number of grains is individually used for training and testing. With increasing number of grains, the error of the prediction is increasing as well, which is shown in \tref{res:complexity}. A large number of grains in the selections increases the systems complexity with many possibilities of magnetization reversal. Here most likely the number of training data has to be extended, in order to improve the prediction accuracy.

\begin{table}[H]
\begin{center}
\begin{tabular}{| c | c | c |}
  \hline
 \bf{\# grains} & \bf{\# selections} & \bf{MAE [T]}\\
    \hline 
    \ \ 1 & 75 & 0.046 \\
     \hline 
    \ \ 2 & 845 & 0.109 \\
     \hline
    \ \ 3 & 378 & 0.143 \\
     \hline
    \ \ 4 & 164 & 0.157 \\
     \hline
    $>$5 & 117 & 0.224 \\
     \hline
\end{tabular}
\end{center}
 \caption{Prediction accuracy according to the number of grains in the selections. PIXEL, GRAIN, SIZE, SW and TWIN are used as training features.}
\label{res:complexity}
\end{table}

\subsection{\label{sec:res_hc}Local coercivity prediction}

\begin{figure}[H]
\begin{center}
 \includegraphics[width=0.7\textwidth]{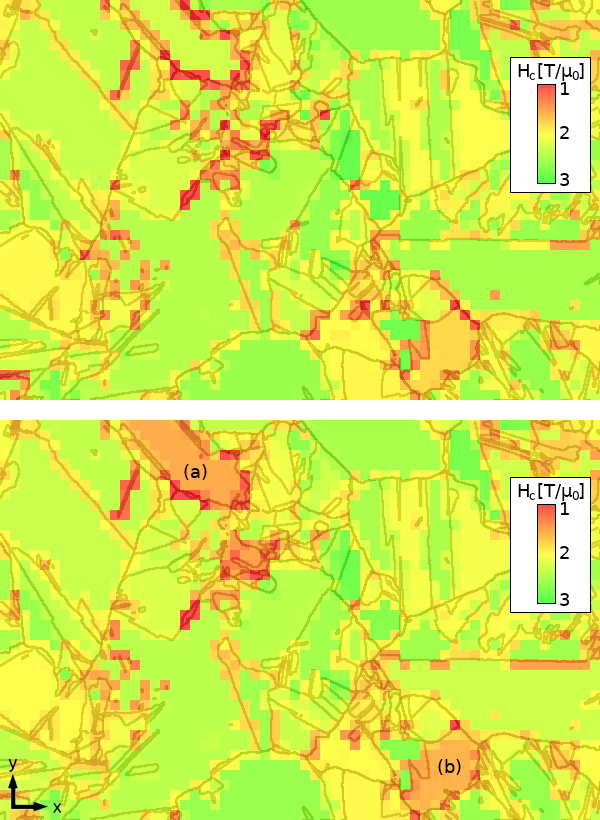}
\end{center}
 \caption{Switching fields $H_\mathrm{c}$ predicted for the EBSD map (600x400 pixels with 0.3 $\upmu$m pixel edge length) by using PIXEL, GRAIN, SIZE, SW, TWIN (top) and GRAIN, SIZE (bottom) as training features.  }
 \label{ebsd_pred}
\end{figure}

In \fref{ebsd_pred} on the top we show coercivity predictions of our decision trees using all 5 features (PIXEL, GRAIN, SIZE, SW, TWIN). Local coercivity is nicely reconstructed with lower values at grain boundaries and higher values for bulk as shown in the simulations (\fref{ebsd} on the bottom). For visual comparison to the previous image we are using only GRAIN and SIZE for predicting the local coercivity of the same EBSD map (\fref{ebsd_pred} on the bottom). The grain boundaries show similar predicted values, whereas two larger grains (a, b) have large deviations. If we imagine the coupling of the local switching fields, it is more likely that the bottom image is correct as nucleation mostly starts at grain boundaries~\cite{Fischbacher2017c}. Due to the very similar prediction results, a simplified demagnetization curve of the whole EBSD map is showing a similar trend as well (see \fref{bulk} in Summary and Conclusion). 

\subsection{\label{sec:res_resolution}Different resolution of EBSD maps}

\begin{figure}[h]
\begin{center}
 \includegraphics[width=0.5\textwidth]{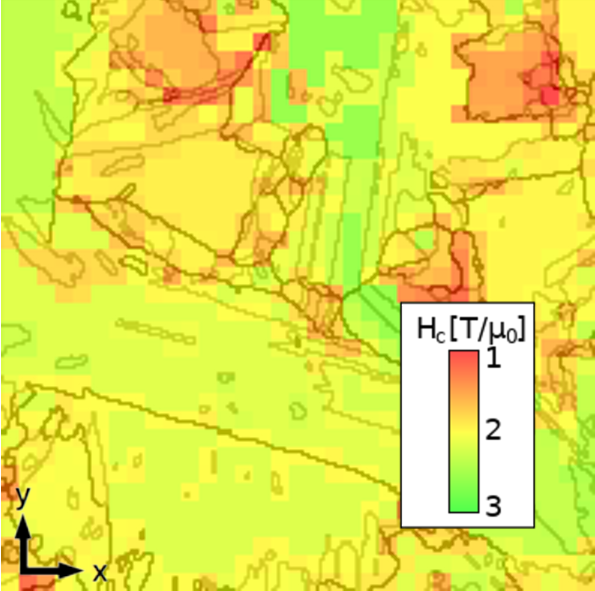}
\end{center}
 \caption{Switching fields $H_\mathrm{c}$ predicted for a new EBSD map with different pixel resolution (200x200 pixels with 0.5 $\upmu$m pixel edge length). The features GRAIN and SIZE are used for training. }
 \label{ebsd_pred3}
\end{figure}

When changing the resolution of an EBSD map we are facing problems using PIXEL for training the model, because the number of data points in the selection is changed. We propose to omit PIXEL and to use the accurately predicting features GRAIN and SIZE to overcome this issue. The physical size of the selection is required to stay about the same, otherwise we cannot be sure that the training data is still valid. \fref{ebsd_pred3} shows the predicted  $H_\mathrm{c}$ for a new EBSD map with a coarser resolution of 0.5 $\upmu$m per pixel in contrast to the resolution of 0.3 $\upmu$m per pixel used for training. The same physical size of the selections is used. 

\subsection{\label{sec:res_prediction}Prediction accuracy of new EBSD maps}

\begin{figure}[H]
\begin{center}
 \includegraphics[width=0.7\textwidth]{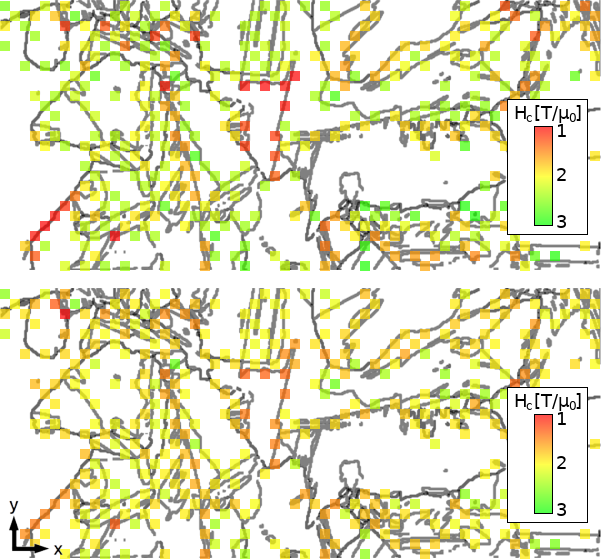}
\end{center}
 \caption{Switching fields $H_\mathrm{c}$ computed (top) and predicted by using the training data from the EBSD map in \fref{ebsd} with the features PIXEL, GRAIN, SIZE, SW, TWIN (bottom) for the EBSD map (600x270 pixels with 0.3 $\upmu$m pixel edge length).  }
 \label{ebsd3}
\end{figure}

With an already trained machine learning model, any new EBSD map from the same material can be predicted in a matter of seconds rather than weeks as needed for micromagnetic simulations. We use a new EBSD map to get the prediction accuracy from the trained trees of the original training data in \fref{ebsd}. Firstly, the machine learning model is trained with the data from \fref{ebsd}. Afterwards the machine learning model is tested with an entirely new dataset from the same MnAl-C sample as in \fref{ebsd} with a size of 180x81 $\upmu$m$^2$ (600x270 pixels). For this new data we compare micromagnetic simulations with the predictions of the previous machine learning model. This time the selections are picked on a regular grid with a size of 10x10 pixels. Each selection with an odd grid number is used for creating a finite element mesh for the simulations. And again the selections are checked for uniqueness to avoid a biased dataset. \fref{ebsd3} on the top shows the computed switching fields $H_\mathrm{c}$ of the new EBSD map from about 400 simulations.

Now we can predict the switching fields on the very same selections by using the trained regression trees from the original EBSD map (\fref{ebsd}). We use a combination of all features: PIXEL, GRAIN, SIZE, SW and TWIN.  \fref{ebsd3} on the bottom highlights those predicted switching fields $H_\mathrm{c}$, which show less maxima and minima compared to the simulated ones. Consequently the MAE is larger than in previous results with a value of 0.23 T. Here again the microstructural complexity, which is discussed in Section~\ref{sec:res_complex}, is probably causing the worsening. More training data need to be produced to overcome this issue.

\section{\label{sec:summary}Summary and Conclusion}

We have demonstrated a fast way to predict local coercivities for MnAl-C. An automated script scans a large EBSD dataset, selects hundreds of unique samples and creates finite element meshes accordingly. For each selection the local switching field $H_\mathrm{c}$ is computed by a fast Landau-Lifshitz-Gilbert micromagnetic solver. The microstructural features of the EBSD selections and its switching fields are used to train Random Forest and Gradient Boosting regressors. A Voting Regressor combines those two methods for an improved prediction accuracy.

Feature selection for the given micromagnetic problem is not straight forward. We are using features like PIXEL, that are a set of multiple features, the data points of the EBSD map. On one hand, training the machine with PIXEL includes the orientation of the grains, the geometry as well as the location information. On the other hand, quite a lot of redundant information is in the dataset. The GRAIN feature includes only the orientation of the single grains and has no geometrical nor location information. This might be the reason that none of the single or combined features reach a MAE value lower than 0.1.

The trained machine can be used to predict local switching fields of new microscopic EBSD data without the need to run new simulations. The time to solution is drastically reduced to seconds instead of days. Even for differing resolutions of the EBSD datasets the training data can be used, because the geometrical information is not absolutely necessary. Due to the microstructural complexity the mean absolute error is higher for the prediction of new EBSD datasets. Here a large number of training data from different EBSD maps need to be produced to improve the prediction accuracies. 

\begin{figure}[H]
\begin{center}
 \includegraphics[width=0.8\textwidth]{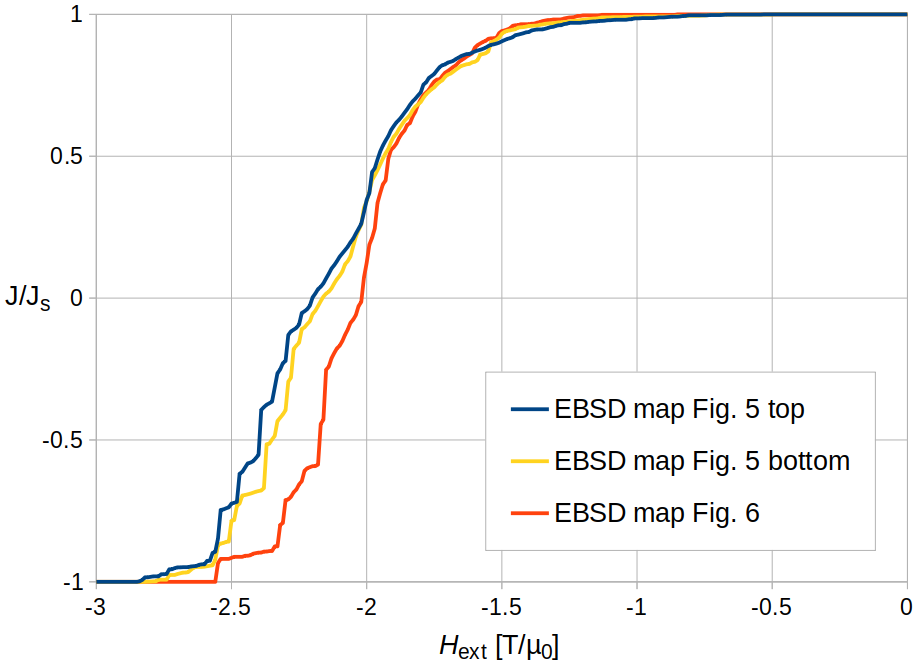}
\end{center}
 \caption{Hysteresis curves of the EBSD maps in \fref{ebsd_pred} and \fref{ebsd_pred3} neglecting stray fields and exchange coupling between selections.}
 \label{bulk}
\end{figure}

Finally we can take all local switching fields from an EBSD map to imitate a demagnetization process and construct a hysteresis curve. We take an increasing external field and as soon as the field value is greater than a locally predicted switching field, the particular selection is magnetically reversed. The ratio from switched to unswitched selections versus the external field gives the simplified demagnetization curve (\fref{bulk}). But we neglect stray field and exchange coupling between the selections. 

In the future we hope to first include the missing interaction contributions between the selections and extend this approach to compute coercivity for bulk MnAl-C. 

\small{\textbf{Acknowledgment} The authors gratefully acknowledge the financial support of the Austrian Science Fund (FWF), Project: I 3288-N36 and the German Research Foundation (DFG), Project: 326646134.\\

\small{\textbf{Data Availability} All data generated or analyzed during this study are available from the corresponding author upon request.\\











\end{document}